\pdfoutput=1
\documentclass[superscriptaddress, amsfonts, amssymb, amsmath, reprint, twoside, nofootinbib]{revtex4-2}

\usepackage[T1]{fontenc}
\usepackage[utf8]{inputenc}

\usepackage{amsthm, mathrsfs}
\usepackage{multirow}
\usepackage{graphicx}
\usepackage[dvipsnames]{xcolor}
\usepackage[normalem]{ulem} %
\usepackage{microtype}

\usepackage[colorlinks,urlcolor=NavyBlue,citecolor=ForestGreen,linkcolor=NavyBlue,pdfusetitle]{hyperref} %
\usepackage[all]{hypcap}
\usepackage{orcidlink}

\graphicspath{{figs/}}

\newcommand{\figQuadScaling}{%
  \begin{figure*}[!t]\vspace{-1em}
    \centering \includegraphics[width=\textwidth]{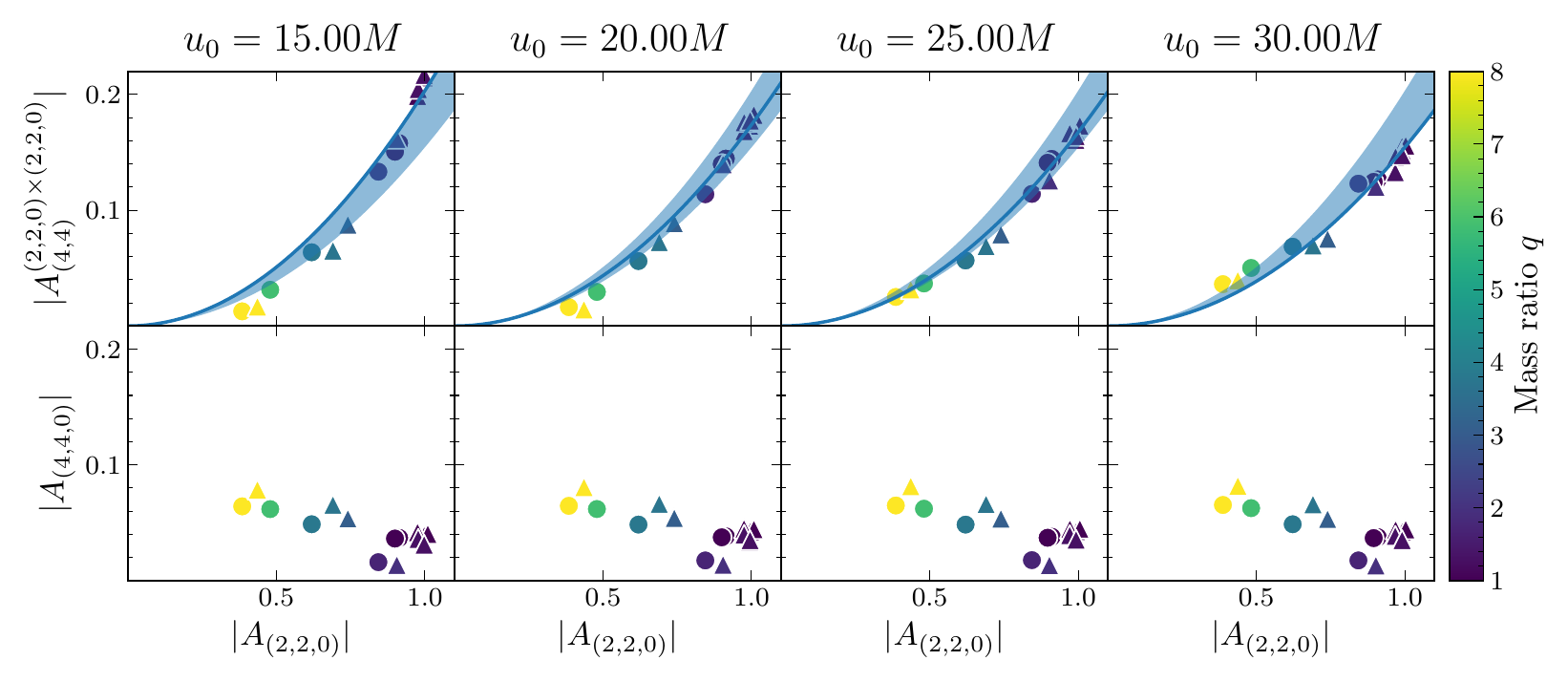}
    \caption{\label{fig:Quad_Scaling}%
      Relationship between the peak amplitudes of the linear $(2,2,0)$ and the quadratic
      $(2,2,0)\times(2,2,0)$ QNMs (top) as well as the linear
      $(4,4,0)$ QNM (bottom), at different model start times
      $u_0$.
      Colors show different mass ratios $q$, and circles and triangles
      denote systems with remnant dimensionless spin
      $\chi_f\approx 0.5$ and $\chi_f \approx 0.7$, respectively.
      Each blue curve is a pure quadratic fit with start time $u_0$,
      and the shaded region brackets every one of the individual fits.\vspace{-1em}}
  \end{figure*}%
}

\newcommand{\figMismatches}{%
  \begin{figure}[tb]
    \centering \includegraphics[width=0.5\textwidth]{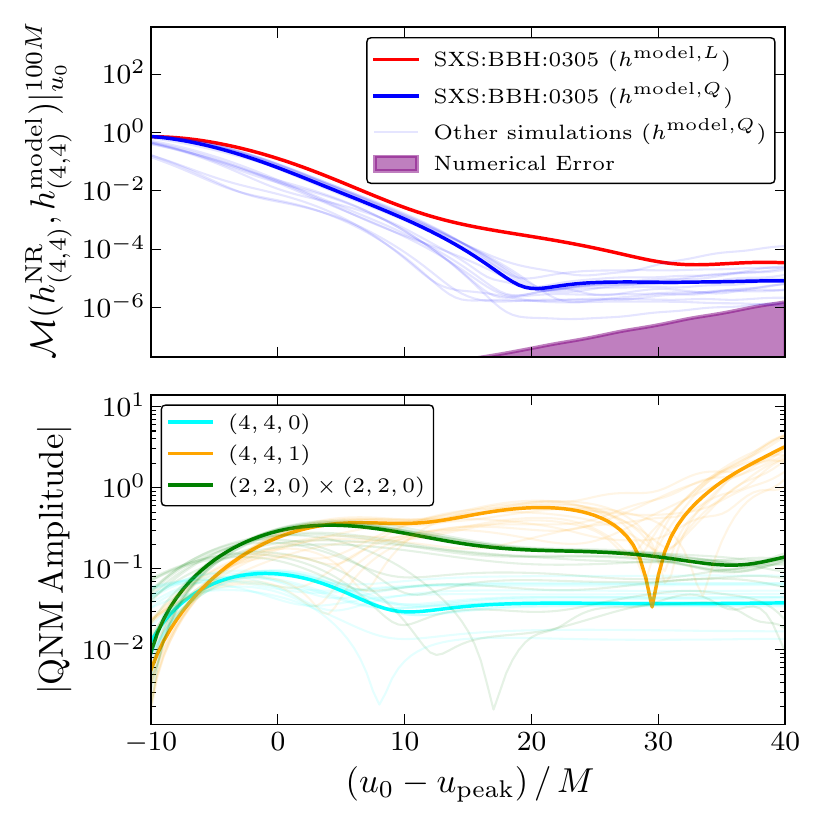}
    \vspace{-2em}
    \caption{\label{fig:mismatch}%
      Top: mismatch in the $(4,4)$ mode for SXS:BBH:0305, as well as for every other simulation examined, and a comparison to the numerical error floor. Bottom: amplitudes of the three QNM terms in the quadratic $(4,4)$ QNM model as a function of the model start time $u_0$.\vspace{-1em}
    }
  \end{figure}%
}

\newcommand{\figResiduals}{%
  \begin{figure}[tb]
    \centering \includegraphics[width=0.5\textwidth]{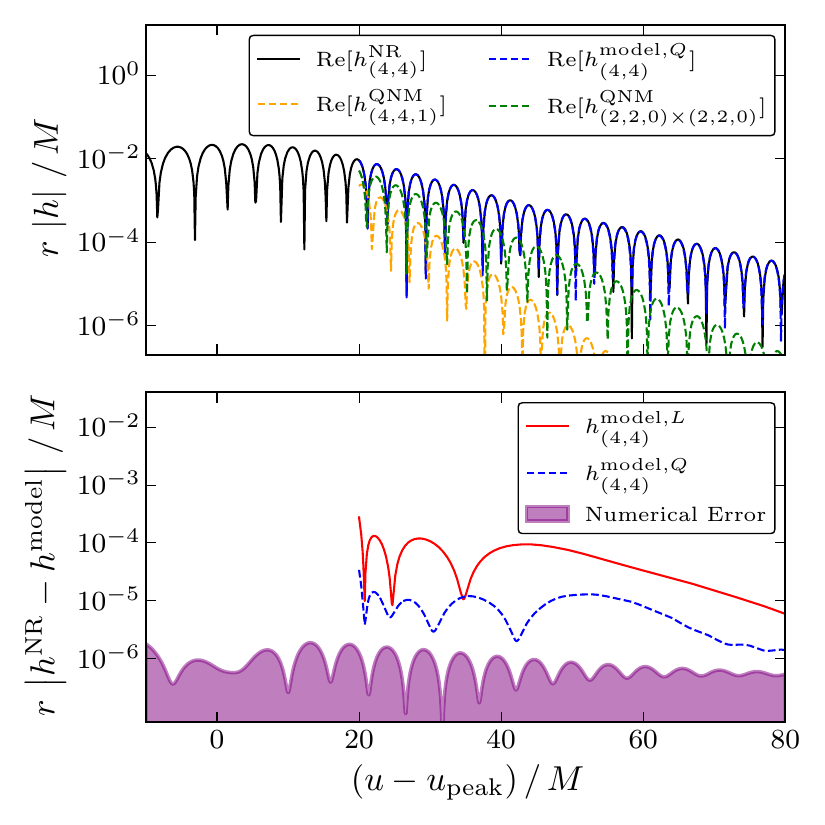}
    \vspace{-2em}
    \caption{\label{fig:residuals}%
     Top: in black, the NR waveform for the SXS:BBH:0305 simulation and its comparison to the quadratic $(4,4)$ QNM model with start time $u_0=20M$
     (total is dashed blue; yellow and green are contributions from individual QNMs, respectively the linear $(4,4,1)$ and the quadratic $(2,2,0)\times(2,2,0)$). 
     Bottom:
     residual in the $(4,4)$ mode when using the linear (solid red) or the quadratic (dashed blue) $(4,4)$
     model. We also show a conservative estimate of the numerical
     error.\vspace{-1em}%
    }
  \end{figure}%
}

\newcommand{\figFrequencies}{%
  \begin{figure}[tb]
    \centering \includegraphics[width=0.5\textwidth]{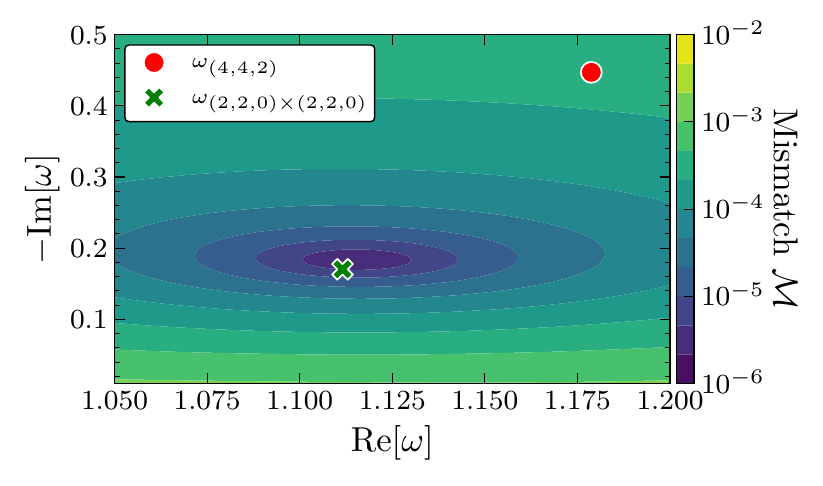}
    \vspace{-2em}
    \caption{\label{fig:frequency}%
      Contour plot of the mismatch between the
      SXS:BBH:0305 waveform and a $(4,4)$ model with three QNMs, in which two
      frequencies are fixed to the GR predictions of the linear $(4,4,0)$
      and $(4,4,1)$ QNMs, but the third is varied. The contour lines are logarithmically spaced in $\mathcal{M}$ between $10^{-6}$ and $10^{-2}$. The start time
      of the model is taken to be $u_{0}=20M$.\vspace{-1em}%
    }
  \end{figure}%
}

\newcommand{\tabSims}{%
\begin{table}[tb]
\centering\vspace{-1em}
\caption{\label{table:sims}%
List of simulations used (ID is shorthand for SXS:BBH:ID from the SXS catalog~\cite{Boyle:2019kee} where the full list of binary parameters can be found) with their mass ratios $q$ and dimensionless remnant spins $\chi_f$. All of these binaries are nonprecessing and are in quasicircular orbits.}
\begin{tabular}{r|cccccccccc} 
\hline\hline\noalign{\smallskip}
ID\ & 1502 & 1476 & 1506 & 1508 & 1474 & 1505 & 1504 & 1485 & 1486 & 1441 \\
\hline
$q$ & $1.00$ & $1.00$ & $1.00$ & $1.28$ & $1.28$  & $1.33$ & $1.98$ & $3.09$ &$3.72$ & $8.00$\\
$\chi_f$ & 0.73 & 0.68 & 0.71 & 0.73 & 0.73 & 0.71 & 0.71 &  0.68 & 0.70 &0.72 \\
\noalign{\smallskip}
ID\  & 1500 & 1492 & 1465 & 1458 & 1438 & 1430 && \multicolumn{1}{r|}{ID} & 0305\\
\cline{1-7}\cline{9-10}
$q$ &  $1.00$ & $1.00$ & $1.71$ & $3.80$ & $5.87$ & $8.00$ && \multicolumn{1}{r|}{$q$} & 1.22 \\
$\chi_f$ & 0.53 & 0.48 & 0.48 &0.47 & 0.47 & 0.50 && \multicolumn{1}{r|}{$\chi_f$} & 0.69 \\
\noalign{\smallskip}\hline\hline
\end{tabular}
\vspace{-1em}
\end{table}%
}

\begin{document}
\title{Nonlinearities in Black Hole Ringdowns}

\newcommand{\Cornell}{\affiliation{Cornell Center for Astrophysics and Planetary
		Science, Cornell University, Ithaca, New York 14853, USA}}
\newcommand{\Caltech}{\affiliation{Theoretical Astrophysics 350-17, California
		Institute of Technology, Pasadena, California 91125, USA}}
\newcommand{\MaxPlanck}{\affiliation{Max Planck Institute for Gravitational
		Physics (Albert Einstein Institute), Am M{\"u}hlenberg 1, D-14476 Potsdam,
		Germany}}
\newcommand{\OleMiss}{\affiliation{Department of Physics and Astronomy,
		University of Mississippi, University, Mississippi 38677, USA}}

\author{Keefe Mitman\,\orcidlink{0000-0003-0276-3856}}
    \email{kmitman@caltech.edu}%
    \Caltech
\author{Macarena Lagos\,\orcidlink{0000-0003-0234-9970}}
    \email{m.lagos@columbia.edu}%
    \affiliation{Department of Physics and Astronomy, Columbia University, New York, New York 10027, USA}
\author{Leo C. Stein\,\orcidlink{0000-0001-7559-9597}}
    \email{lcstein@olemiss.edu}
    \OleMiss
\author{Sizheng Ma\,\orcidlink{0000-0002-4645-453X}}
    \Caltech
\author{Lam Hui}
    \affiliation{Department of Physics and Astronomy, Columbia University, New York, NY 10027, USA}
\author{Yanbei Chen\,\orcidlink{0000-0002-9730-9463}}
    \Caltech
\author{\\Nils Deppe\,\orcidlink{0000-0003-4557-4115}} \Caltech
\author{Fran\c{c}ois H\'{e}bert\,\orcidlink{0000-0001-9009-6955}}\Caltech
\author{Lawrence E. Kidder\,\orcidlink{0000-0001-5392-7342}} \Cornell
\author{Jordan Moxon\,\orcidlink{0000-0001-9891-8677}} \Caltech
\author{Mark A. Scheel\,\orcidlink{0000-0001-6656-9134}}
\Caltech
\author{Saul A. Teukolsky\,\orcidlink{0000-0001-9765-4526}}
\Caltech
\Cornell
\author{William Throwe\,\orcidlink{0000-0001-5059-4378}} \Cornell
\author{Nils L. Vu\,\orcidlink{0000-0002-5767-3949}} \MaxPlanck

\date{\today} %

\begin{abstract}
  The gravitational wave strain emitted by a perturbed black hole (BH)
  ringing down is typically modeled analytically using first-order BH
  perturbation theory.
  In this Letter we show that second-order effects are necessary for
  modeling ringdowns from BH merger simulations.
  Focusing on the strain's $(\ell,m)=(4,4)$ angular harmonic, we show the presence of a quadratic effect across a range of
  binary BH mass ratios that agrees with theoretical expectations.
  We find that the quadratic $(4,4)$ mode's
  amplitude exhibits quadratic scaling with the fundamental $(2,2)$ mode---its parent mode. The nonlinear mode's amplitude is comparable to or even larger than that of the linear $(4,4)$ mode.
  Therefore, correctly modeling the ringdown of higher harmonics---improving mode mismatches by up to 2 orders of magnitude---requires the inclusion of nonlinear effects.
\end{abstract}

\maketitle

\figQuadScaling

Nonlinearity is responsible for the rich phenomenology of general
relativity (GR). While many exact nonlinear solutions are
known~\cite{Stephani:2003tm,Griffiths:2009dfa},
LIGO-Virgo-KAGRA observables---gravitational waves (GWs) from merging binary black holes
(BHs)---must be predicted by numerical relativity (NR). Analytic
perturbation theory has an important role far from the merger: at
early times, post-Newtonian (PN) theory, and at late times (ringdown),
black hole perturbation theory~\cite{Regge:1957td,
  Zerilli:1970wzz, Teukolsky:1973ha}, provided that the remnant asymptotes
to a perturbed Kerr BH~\cite{Penrose:1969pc, Chrusciel:2012jk}. PN
theory has been pushed to high perturbative
order~\cite{Blanchet:2013haa}, but the standard paradigm for modeling
ringdown is only linear theory (see~\cite{Berti:2009kk} for a review).
It may then come as a surprise if linear theory can be used to model
ringdown even at the peak of the
strain~\cite{Giesler:2019uxc, Bhagwat:2019dtm, Cook:2020otn, JimenezForteza:2020cve, Dhani:2020nik,Finch:2022ynt}, the most nonlinear phase of a
BH merger.

The ``magic''
nature of the Kerr geometry~\cite{Teukolsky:2014vca} leads to a
decoupled, separable wave equation for first-order
perturbations (the Teukolsky equation~\cite{Teukolsky:1973ha}),
schematically written as
\begin{align}
  \label{eq:teuk-sketch}
  \mathcal{T} \psi = \mathcal{S},
\end{align}
where $\mathcal{S}$ is a source term that vanishes for linear perturbations in vacuum, $\psi$ is related to the first-order
correction to the curvature scalar $\psi_{4}$, and the linear differential Teukolsky
operator $\mathcal{T}$ depends on the dimensionless spin parameter
$\chi\equiv |S|/M^{2}$ through the combination $a=|S|/M$, where $S$ is the BH spin angular momentum and $M$ is the BH mass (throughout we
use geometric units $G=c=1$).
The causal Green's function $\mathcal{G}\sim\mathcal{T}^{-1}$ has an infinite, but discrete
set of complex frequency poles $\omega_{(\ell, m, n)}$.\footnote{%
For this study we focus only on prograde modes (in the sense described
in~\cite{MaganaZertuche:2021syq}), and therefore omit the additional
prograde/retrograde label $\pm$. The Green's function also has branch
cuts, which lead to power-law tails~\cite{Leaver:1986gd}, which we
ignore here.}
This makes GWs
during ringdown well described by a superposition of exponentially
damped sinusoids, called quasinormal modes (QNMs). The real and
imaginary parts of $\omega_{(\ell, m, n)}$ determine the QNM
oscillation frequency and decay timescale, respectively. These modes
are labeled by two angular harmonic numbers $(\ell,m)$ and an overtone
number $n$. The
combination $M\omega_{(\ell, m, n)}$ is entirely determined by $\chi$.

To date, the linear QNM spectrum has been used to analyze current GW
detections~\cite{Isi:2019aib, Finch:2022ynt, Cotesta:2022pci, Isi:2022mhy},
forecast the future detectability of ringdown~\cite{Berti:2016lat,
  Ota:2019bzl, Bhagwat:2021kwv}, and perform tests of gravity in the
strong field regime~\cite{Berti:2018vdi, LIGOScientific:2021sio}.

Since the sensitivity of GW detectors will increase in the coming years~\cite{KAGRA:2013rdx, LISA, Maggiore:2019uih, Evans:2021gyd}, there is the potential to observe
nonlinear ringdown effects in high signal-to-noise ratio (SNR) events.
A few previous works have shown that second-order perturbation effects
can be identified in some NR simulations of binary BH
mergers~\cite{London:2014cma, Ma:2022wpv}.
In this Letter we show that quadratic QNMs---the
damped sinusoids coming from second-order perturbation theory in
GR---are a ubiquitous effect present in simulations across various
binary mass ratios and remnant BH spins. In particular, for
the angular harmonic $(\ell,m)=(4,4)$, we find that the quadratic QNM
amplitude exhibits the expected quadratic scaling relative to its parent---the fundamental $(2,2)$ mode. The quadratic amplitude also has a value that is comparable to that of the linear $(4,4)$ QNMs for every
simulation considered, thus highlighting the need to include
nonlinear effects in ringdown models of higher harmonics.

\textit{Quadratic QNMs.}---%
Second-order perturbation theory has been studied for both Schwarzschild and Kerr BHs~\cite{Gleiser:1996yc,
  Gleiser:1995gx, Gleiser:1998rw, Ioka:2007ak, Nakano:2007cj,
  Okuzumi:2008ej, Brizuela:2009qd, Pazos:2010xf, Ripley:2020xby,
  Loutrel:2020wbw, Lagos:2022otp}. This involves the same Teukolsky operator as in
Eq.~\eqref{eq:teuk-sketch} acting on the second-order curvature correction, and a complicated source $\mathcal{S}$ that depends quadratically on the linear perturbations~\cite{Campanelli:1998jv, Loutrel:2020wbw, Ripley:2020xby}.
The second-order solution results from a rather involved integral of
this source against the Green's function $\mathcal{G}$~\cite{Okuzumi:2008ej,Lagos:2022otp}. We only
need to know that it is quadratic in the linear perturbation and that,
after enough time, it is well approximated by the quadratic QNMs.

The frequency spectrum of quadratic QNMs is distinct from the linear
QNM spectrum. For each pair of linear QNM frequencies
$\omega_{(\ell_{1},m_{1},n_{1})}$ and
$\omega_{(\ell_{2},m_{2},n_{2})}$ (in either the left or right half complex plane), there will be a corresponding quadratic QNM frequency
\begin{align}
    \omega&\equiv\omega_{(\ell_{1},m_{1},n_{1})}+\omega_{(\ell_{2},m_{2},n_{2})}.
\end{align}
As the linear $(2,\pm 2,0)$ modes are most important, it is promising
to investigate the quadratic QNMs they generate, which primarily appear in the $(\ell, m)=(4,\pm4)$ modes~\cite{Ioka:2007ak, Nakano:2007cj, Lagos:2022otp}. The quadratic
QNM coming from the $(2,2)$ mode would have frequency $\omega_{(2,2,0)\times
  (2,2,0)}\equiv2\omega_{(2,2,0)}$ and would decay
faster than the linear fundamental mode $(4,4,0)$, but slower
than the first linear overtone $(4,4,1)$, regardless of the
BH spin.\footnote{%
  The $(\ell, m,n)=(2,2,0)$ can excite other quadratic QNMs with
  frequency $\omega=\omega_{(2,2,0)}-\overline{\omega_{(2,2,0)}}$. These will
  instead be related to the memory effect, as they are
  non-oscillatory. From angular selection rules they will be most
  prominent in the $(2,0)$ mode. While these effects could also prove interesting to study, they are much more well understood than the quadratic QNMs in the $(4,4)$ mode, so we reserve their examination for future work~\cite{Mitman:2020pbt,MaganaZertuche:2021syq}.}

The NR strain at future null infinity contains all of the angular information of the GW and is decomposed as
\begin{align}
h^{\text{NR}}(u,\theta,\phi)\equiv\sum\limits_{\ell=2}^{\infty}\sum\limits_{|m|\leq\ell}h_{(\ell, m)}^{\text{NR}}(u)\phantom{}_{-2}Y_{(\ell,m)}(\theta,\phi),\label{hNR}
\end{align}
where $u$ is the Bondi time and ${}_{-2}Y_{(\ell,m)}$ are the spin-weighted $s=-2$ spherical harmonics.
We model this data with two different QNM Ansätze, valid between times $u \in [u_0, u_f]$.
The first model, which is typically used in the literature,
involves purely linear QNMs,
\begin{align}
  \label{hL}
  h_{(\ell,m,N)}^{\text{model},\,L}(u)&=\sum_{n=0}^{N} A_{(\ell,m,n)} e^{-i \omega_{(\ell,m,n)} (u-u_\text{peak})}.
\end{align}
Here $A_{(\ell,m,n)}$ is the peak amplitude of the linear QNM with frequency $\omega_{(\ell,m,n)}$, $N$ is the total number
of overtones considered in the model, and $u_\text{peak}$ is the time at which the $L^{2}$ norm of the
strain over the two-sphere achieves its maximum value (a proxy for the
merger time), which we take to be $u_\text{peak}=0$ without loss of generality. Note that here we have suppressed
the spheroidal-spherical decomposition (which we include as in Eq.\ (6)
of~\cite{MaganaZertuche:2021syq}).

We will use Eq.~\eqref{hL} to model both the $(2,2)$ and $(4,4)$
modes of the strain.\footnote{We ignore the $m<0$ modes because the binary BH simulations that we consider are nonprecessing and are in quasicircular orbits, so the $m<0$ modes can be recovered from the $m>0$ modes via $h_{(\ell,m)}=(-1)^{\ell}\overline{h_{(\ell,-m)}}$} When modeling the $(2,2)$ mode, we use $N=1$ and
when modeling the $(4,4)$ mode we use $N=2$. While prior works have
included more overtones in their models~\cite{Giesler:2019uxc,Bhagwat:2019dtm, Cook:2020otn, JimenezForteza:2020cve, Dhani:2020nik, MaganaZertuche:2021syq}, we restrict ourselves to no more than two
overtones because we find that the amplitudes of higher overtones tend to
vary with the model start time $u_{0}$ and hence are not very robust. Moreover, their inclusion does not affect considerably the best-fit amplitude of the modes in which we are interested.

The novel QNM model, which includes second-order effects and highlights our main result, only changes how
the $(4,4)$ mode is described, compared to Eq.~\eqref{hL}.  It is given by
\begin{align}
    \label{hQ}
    h^{\text{model},\,Q}_{(4,4)}(u) &=  \sum_{n=0}^{1}  A_{(4,4,n)} e^{-i \omega_{(4,4,n)} (u-u_\text{peak})}\\
    &\phantom{=.}+ A^{(2,2,0)\times (2,2,0)}_{(4,4)} e^{-i \omega_{(2,2,0)\times (2,2,0)} (u-u_\text{peak})},
    \nonumber
\end{align}
where $A_{(4,4)}^{(2,2,0)\times (2,2,0)}$ is the peak amplitude of the
quadratic QNM sourced by the linear $(2,2,0)$ QNM interacting
with itself.
In each model, for the linear amplitudes we factor out the angular mixing coefficients, whereas for the quadratic term we absorb the angular structure (from the nonlinear mixing coefficients and the Green's function integral of the second-order source terms) into the amplitude $A_{(4,4)}^{(2,2,0)\times (2,2,0)}$.
We emphasize that the two models
$h_{(4,4,2)}^{\text{model},\,L}(u)$ and
$h^{\text{model},\,Q}_{(4,4)}(u)$ contain the same number of free
parameters.

In these ringdown models, we fix the QNM frequencies to the values
predicted by GR in vacuum and fit the QNM amplitudes to NR simulations, which
cannot be predicted from first principles as they depend on the
merger details. From the quadratic sourcing
by the linear $(2,2,0)$ mode, we expect
$A^{(2,2,0)\times (2,2,0)}_{(4,4)}\propto (A_{(2,2,0)})^2$. We will
use this theoretical expectation as one main test to confirm the
presence of quadratic QNMs. To perform this check we need a family of
systems with different linear amplitudes, which is easily accomplished
by varying the binary mass ratio $q\equiv m_{1}/m_{2}\ge 1$.

The proportionality coefficient between $(A_{(2,2,0)})^2$ and
$A^{(2,2,0)\times (2,2,0)}_{(4,4)}$ (which we expect to be order
unity~\cite{London:2014cma, Lagos:2022otp}) comes from the spacetime
dependence of the full quadratic source as well as the Green's function.
While, in principle, this can be computed, we use the fact that it
should only depend on the dimensionless spin $\chi_{f}$ of the remnant
BH.

\tabSims

We consider a family of 17 simulations (listed in
Table~\ref{table:sims}) of binary BH systems in the range $q\in
[1,8]$. To control the dependence on $\chi_{f}$, six are in the range
$\chi_f=0.5\pm 0.035$, and ten have $\chi_f=0.7\pm 0.035$. The final simulation, SXS:BBH:0305, is consistent with GW150914~\cite{LIGOScientific:2016vbw}. These simulations were produced using the Spectral Einstein Code (\texttt{SpEC}) and are available in the SXS catalog~\cite{SpECCode,SXSCatalog,Boyle:2019kee}. For each simulation, the strain waveform has been extracted using Cauchy characteristic extraction and has then been mapped to the superrest frame at $250M$ after $u_\text{peak}$~\cite{Moxon:2020gha,Moxon:2021gbv,CodeSpECTRE,Mitman:2021xkq,Mitman:2022kwt} using the techniques presented in~\cite{Mitman:2022kwt} and the code \texttt{scri}~\cite{scri_url,Boyle:2013nka,Boyle:2014ioa,Boyle:2015nqa}.

\textit{Quadratic fitting.}---%
In order to fit the ringdown models to the NR waveforms, using the least-squares implementation from SciPy v1.6.2~\cite{2020SciPy-NMeth}, we minimize the $L^{2}$ norm of the residual
\begin{align}
    \langle R, R\rangle\quad\text{for}\quad R\equiv h_{(\ell,m)}^{\text{NR}}-h_{(\ell,m)}^{\text{model}}, \label{eq:res}
\end{align}
where the inner product between modes $a$ and $b$ is
\begin{equation}
    \langle a,b\rangle\equiv\int_{u_{0}}^{u_{f}}du\,\overline{a(u)}b(u),
\end{equation}
with $\overline{a(u)}$ being the complex conjugate of $a(u)$. We will fix $u_f=100M$ and vary the value of $u_0$. In Eq.~\eqref{eq:res}, $h^{\text{model}}$ is given by Eq.~\eqref{hL} with $N=1$ for the $(2,2)$ mode and Eq.~\eqref{hQ} for the $(4,4)$ mode by default, unless explicitly mentioned that we use the purely linear model, Eq.~\eqref{hL}, with $N=2$.
We fix the frequencies and perform a spheroidal-to-spherical angular decomposition of the linear terms in our QNM models using the open-source \texttt{Python} package \texttt{qnm}~\cite{Stein:2019mop}.

We show the main result of the fits in Fig.~\ref{fig:Quad_Scaling} for a range of initial times $u_0$ with which we find the best-fit amplitudes to be stable (shown later). In the top panel, we see that $A_{(2,2,0)}$ and $A_{(4,4)}^{(2,2,0)\times (2,2,0)}$ are consistent with a quadratic relationship, illustrated by the shaded blue region that is obtained by combining the fitted quadratic curves for $u_0\in[15M,30M]$. In this region, we find the ratio $A_{(4,4)}^{(2,2,0)\times (2,2,0)}/(A_{(2,2,0)})^2$ to range between 0.20 and 0.15.\footnote{%
In addition to the amplitudes, we can also check the consistency of the phases of the quadratic $(4,4)$ QNM and the linear $(2,2,0)$ QNM. We find that the phase of $A_{(4,4)}^{(2,2,0)\times (2,2,0)}/A_{(2,2,0)}^2$ is always within 0.4 radians of 0, for each simulation, for start times in the range $u_{0}\in[15M,30M]$.} Again we emphasize that here $A_{(2,2,0)}$ has the mixing coefficients factored out, while $A_{(4,4)}^{(2,2,0)\times (2,2,0)}$ contains whatever angular structure arises through nonlinear effects.
There is no noticeable difference in the quadratic relationship followed by the $0.7$ and $0.5$ spin families of waveforms, compared to the variations that are observed in the best-fit $A^{(2,2,0)\times (2,2,0)}_{(4,4)}$ due to the choice of the model start time $u_0$.

We emphasize that this quadratic behavior is unique to the $A_{(4,4)}^{(2,2,0)\times (2,2,0)}$ mode, as can be seen in the bottom panel of Fig.~\ref{fig:Quad_Scaling}, where we show the best-fit linear amplitude $A_{(4,4,0)}$ as a function of $A_{(2,2,0)}$. These two modes are not related quadratically (for more on their scaling with mass ratio, see~\cite{Borhanian:2019kxt}), which confirms the distinct physical origin of $A_{(4,4,0)}$ and $A^{(2,2,0)\times (2,2,0)}_{(4,4)}$.  The best-fit amplitudes of  $A_{(4,4,0)}$ and $A_{(2,2,0)}$ are nearly constant across these values of $u_0$, which is why the four bottom figures look the same.
A key result of Fig.~\ref{fig:Quad_Scaling} is that $A^{(2,2,0)\times (2,2,0)}_{(4,4)}$ is comparable to or larger (by a factor of $\sim 4$ in cases with $q \approx 1$) than $A_{(4,4,0)}$ at the time of the peak. Given that the exponential decay rates of $A^{(2,2,0)\times (2,2,0)}_{(4,4)}$ and $A_{(4,4,0)}$ for a BH with $\chi_{f} = 0.7$ are $\text{Im}[M\omega_{(2,2,0)\times (2,2,0)}]=-0.16$ and $\text{Im}[M\omega_{(4,4,0)}]=-0.08$, respectively, even beyond $10 M$ after $u_\text{peak}$ the quadratic mode will be larger than the linear mode for equal mass ratio binaries.\footnote{We also find the peak amplitude $A_{(4,4,1)}$ to be comparable or sometimes larger than $A^{(2,2,0)\times (2,2,0)}_{(4,4)}$ (see bottom panel of Fig.~\ref{fig:mismatch}) but, since $\text{Im}[M\omega_{(4,4,1)}]=-0.25$, this $(4,4,1)$ mode decays fast enough that it will be comparable or smaller than the quadratic $(4,4)$ mode after $u=10M$.}
Thus, for large SNR events in which the $(4,4)$ mode is detectable, the quadratic QNM could be measurable. 

\textit{Comparisons.}---%
Figure~\ref{fig:residuals} shows the GW150914 simulation (SXS:BBH:0305) and its fitting at $u_0=20M$, the time at which the residual in the $(4,4)$ mode reaches its minimum. The top panel shows the waveform fit with the $(4,4)$ quadratic model $h^{\text{model},\,Q}_{(4,4)}$ as a function of time, where we find that it can fit rather well the amplitude and phase evolution of the numerical waveform at late times. 
The bottom panel shows the residual of the NR waveform with the linear and quadratic $(4,4)$ QNM models, $h^{\text{model},\,L}_{(4,4,2)}$ and $h^{\text{model},\,Q}_{(4,4)}$, and a conservative estimate for the numerical error obtained by comparing the highest and second highest resolution simulations for SXS:BBH:0305.
\figResiduals
We see that even though the linear and quadratic $(4,4)$ models have the same number of free parameters, the residual of $h^{\text{model},\,Q}_{(4,4)}$ is nearly an order of magnitude better, which confirms the importance of including quadratic QNMs. Since, in general, the quadratic mode decays in time slower than the $(4,4,2)$ QNM, the quadratic model generally better describes the late time behavior of the waveform. In addition, the best-fit value of $A_{(4,4,0)}$---which is the most important QNM in the $(4,4)$ mode at late times---differs in the linear and quadratic models, which causes the residuals to be rather different even beyond $u=50M$ when we expect the overtones and quadratic mode to be subdominant.

\figMismatches

In addition to the residuals, we quantify the goodness of fit by our models through the mismatch
\begin{equation}
    \label{eq:mismatch}
    \mathcal{M} = 1-\text{Re} \left[\frac{\langle h_{(\ell,m)}^\text{NR} | h_{(\ell,m)}^{\text{model}} \rangle}{\sqrt{\langle h_{(\ell,m)}^\text{NR} | h_{(\ell,m)}^\text{NR} \rangle\langle h_{(\ell,m)}^{\text{model}} | h_{(\ell,m)}^{\text{model}} \rangle}}\right].
\end{equation}
The top panel of Fig.~\ref{fig:mismatch} shows the mismatch in the $(4,4)$ mode between the NR waveform and the QNM model as a function of $u_0$. The red and blue lines show the results for the SXS:BBH:0305 simulation when the $(4,4)$ mode was modeled with $h^{\text{model},\,L}_{(4,4,2)}$ and $h^{\text{model},\,Q}_{(4,4)}$, respectively. As a reference, we also show the numerical error calculated for SXS:BBH:0305.\footnote{The numerical error for the other simulations tends to be worse since they were not run with as fine of a resolution, but the errors are nonetheless comparable to that of SXS:BBH:0305.} We see that the numerical error is below the fitted model mismatches for $u_0\lesssim40M$, but will cause the mismatch to worsen at later times.
We also see that the linear model performs worse than the quadratic model for any $u_0$, confirming that the residual difference shown in the bottom panel of Fig.~\ref{fig:residuals} was not a coincidence of the particular fitting time chosen there. At times $u_0\approx 20M$, we see that the mismatch is about 2 orders of magnitude better in the quadratic model. 
We find similar results for all of the simulations analyzed in this Letter\footnote{Except for a few simulations at early times $0\lesssim u_0\lesssim10M$, for which the linear model can have a marginally better mismatch.} (light blue thin curves show the mismatch of the $h^{\text{model},\,Q}_{(4,4)}$ in those simulations), although the mismatch difference becomes more modest for simulations with $q\approx 8$ since the relative amplitude of the quadratic mode decreases (cf.\ bottom panel of Fig.\ \ref{fig:Quad_Scaling} where we see that amplitude of the $(2,2,0)$ mode decreases with $q$, while the amplitude of the $(4,4,0)$ mode increases with $q$). When comparing the mismatches to the error, we find that every simulation remains above the numerical error floor until $u_{0}\gtrsim 40M$.\footnote{We emphasize that the reason the numerical error curve increases with $u_{0}$ is because of the normalization factor in Eq.~\eqref{eq:mismatch}; i.e., with higher $u_{0}$ the integral of the numerical error becomes more comparable to the strain's amplitude.}

In the bottom panel of Fig.~\ref{fig:mismatch}, we show the best-fit amplitudes of the QNMs in the $(4,4)$ mode as functions of $u_0$. We show the results for SXS:BBH:0305 (thick lines) as well as the rest of the simulations (thin lines). We see that at $u_0\gtrsim10M$ the amplitude of $A_{(4,4,0)}$ is extremely stable, but the faster the additional QNM decays, the more variations that are seen. Nevertheless, the $A^{(2,2,0)\times (2,2,0)}_{(4,4)}$ exhibits only $\sim20\%$ variations for $u_{0}\in[15M,30M]$, whereas $A_{(4,4,1)}$ varies by $\sim90\%$ in the same range. Before and near $u_{0}\approx10M$ every amplitude shows considerable variations, which is why we use $u_0\geq 15M$ in this Letter. This suggests a need to improve the QNM model, either by including more overtones as in~\cite{Giesler:2019uxc}, modifying the time dependence of the linear~\cite{Sberna:2021eui} and quadratic terms, or considering more nonlinear effects.

Finally we check which frequency is preferred by the $(4,4)$ mode of the numerical strain. For this, we fix two frequencies to be the linear $\omega_{(4,4,0)}$
and $\omega_{(4,4,1)}$ frequencies, and keep one frequency free. We vary the frequency of that third
term and fit every amplitude to minimize the residual in Eq.~\eqref{eq:res}. Figure~\ref{fig:frequency} shows contours of the mismatch over the real and
imaginary parts of the unknown frequency for the SXS:BBH:0305 simulation
using $u_0=20M$. We confirm that the data clearly
prefers the frequency $\omega_{(2,2,0)\times (2,2,0)}=2\omega_{(2,2,0)}$ over
$\omega_{(4,4,2)}$.

\figFrequencies

\textit{Conclusions.}---%
We have shown that second-order effects are present
in the ringdown phase of binary BH mergers for a wide range of mass
ratios, matching theoretical expectations and helping improve
ringdown modeling at late times.
We analyzed 17 NR simulations and in every one of them we found that, in the $(\ell,m)=(4,4)$ mode, the quadratic QNM analyzed has a peak amplitude that is comparable to or larger than the $(\ell, m,n)=(4,4,0)$ fundamental linear QNM. Because of the relatively slow decay of this quadratic QNM, we find that for nearly equal-mass systems this QNM will be larger than the corresponding linear fundamental mode even $10M$ after $u_{\text{peak}}$.

These results highlight that we may be able to observe this nonlinear effect in future high-SNR GW events with a detectable $(4,4)$ harmonic.
A quantitative analysis, and a generalization to other harmonics, will be performed in the future to assess in detail the detectability of quadratic QNMs and how well they can be distinguished from linear QNMs, for current GW detectors at design sensitivity as well as next-generation GW detectors. It would also be interesting to study how the linear/quadratic relationship of these nonlinearities varies with the spin of the remnant, especially as one approaches maximal spin. 

The confirmation of quadratic QNMs opens new possibilities for more general understanding of the role of nonlinearities in the ringdown of perturbed black holes. It is now clear that we can readily improve the basic linear models that have been used previously in theoretical
and observational ringdown analyses. Quadratic QNMs provide new opportunities to
maximize the science return of GW detections, by increasing the likelihood of detecting multiple QNM frequencies.
One of these key science goals is performing high-precision consistency tests of GR with GW observations.
Fulfilling this aim will require a correct ringdown model, which incorporates the nonlinear effects that we have shown to be robustly present.

\textit{Acknowledgments.}---%
We thank
Max Isi
and
the Flatiron Institute for fostering discourse,
and
Vishal Baibhav,
Emanuele Berti, Mark Cheung, Matt Giesler, Scott Hughes, and Max Isi
for valuable 
conversations.
Computations for this work were performed with the Wheeler cluster at Caltech. This work was supported in part by the Sherman Fairchild Foundation and by NSF Grants No.~PHY-2011961, No.~PHY-2011968, and No.~OAC-1931266 at Caltech, as well as NSF Grants No.~PHY-1912081, No.\ PHY-2207342, and No.~OAC-1931280 at Cornell.
The work of L.C.S. was partially supported by NSF CAREER Grant No.
PHY-2047382.
M.L. was funded by the Innovative Theoretical Cosmology Fellowship at Columbia University. L.H. was funded by the DOE DE-SC0011941 and a Simons Fellowship in Theoretical Physics.
M.L. and L.C.S. thank the Benasque Science Center and the organizers
of the 2022 workshop ``New frontiers in strong gravity,'' where some
of this work was performed; and M.L. acknowledges NSF Grant
No.~PHY-1759835 for supporting travel to this workshop.

\textit{Note added.}--Recently, we learned that Cheung \textit{et al.} conducted a similar study, whose results are consistent with ours~\cite{Cheung:2022rbm}.

\makeatletter
\def\bibsection{%
   \par
   \begingroup
    \baselineskip26\p@
    \bib@device{\hsize}{72\p@}%
   \endgroup
   \nobreak\@nobreaktrue
   \addvspace{19\p@}%
  }
\makeatother

\bibliography{QNM_Refs}

\end{document}